\documentclass[final,5p,times,twocolumn]{elsarticle}
\usepackage{amsmath,amssymb,amsfonts}
\usepackage{graphicx}
\usepackage{hyperref}
\usepackage{slashed}
\usepackage{booktabs,tabulary}
\usepackage{mciteplus}
\usepackage{color}
\usepackage{physics}

\usepackage{soul}

\usepackage{bibentry}
\newcommand{\ignore}[1]{}
\newcommand{\nobibentry}[1]{{\let\nocite\ignore\bibentry{#1}}}

\makeatletter
\def\bibinfo@X@title#1,{\ignorespaces}
\makeatother

\begin{document}

\begin{frontmatter}

\title{The role of charged exotic states in $e^+e^- \to \psi(2S) \;  \pi^+ \pi ^-$}
\author[Mainz]{Daniel A. S. Molnar}
\author[Mainz]{Igor Danilkin}
\author[Mainz]{Marc Vanderhaeghen}
\address[Mainz]{Institut f\"ur Kernphysik \& PRISMA$^+$  Cluster of Excellence, Johannes Gutenberg Universit\"at,  D-55099 Mainz, Germany}

\begin{abstract}
In this work, we use the dispersion theory to provide a physical description of recent BESIII data on the reaction $ e^+ e^- \to \psi (2S) \, \pi^+ \, \pi^-$. 
Taking into account explicitly the effects of charged exotic intermediate states in the $t$- and $u$-channels as well as the two-pion final state interaction, we describe the invariant mass distribution for four different $e^+ e^-$ center-of-mass energies. The effects of the $\pi\pi$ rescattering are accounted for in a single channel Omn\`es approach which is found to explain the $\pi \pi$-invariant mass distributions at all $e^+ e^-$ center-of-mass energies. For $q= 4.226$ GeV and $q= 4.258$ GeV the already established charged exotic state $Z_c(3900)$ is considered as the intermediate state, whereas for $q= 4.358$ GeV the rescattering of pions dominates the fits. For the highest energy, $q= 4.416$ GeV, a heavier charged exotic state with mass $m_{Z_c} = 4.016(4)$ GeV and width $\Gamma_{Z_c} = 52(10)$ MeV is essential to describe the experimental data. Although the mass of this state is consistent with the established $Z_c(4020)$, its width is significantly larger.

\end{abstract}

\end{frontmatter}

\section{Introduction}\label{intro}

The state $Z_c^-(4430)$ was the first charged charmonium-like state observed in the invariant mass distribution of $B$ decays, $B \to K \pi^- \psi(2S)$, by the Belle Collaboration in 2007  \cite{Choi:2007wga}. Seven years later this state was confirmed by LHCb \cite{Aaij:2014jqa} and observed again by Belle \cite{Chilikin:2013tch}. Simultaneously, the BESIII Collaboration discovered a new charged exotic state $Z_c(3900)$ from electron-positron annihilation $e^- e^+ \to \pi^+ \pi^- J/\psi$ \cite{Ablikim:2013mio}. From there on, more than five new charged states were claimed to be observed experimentally in the charmonium sector \cite{Tanabashi:2018oca,Yuan:2018inv, Olsen:2017bmm}. 

Many mechanisms have been studied to explain the nature of these charged states. Since it is necessary to have at least four quarks to provide the electric charge, extensions of the conventional $q\bar{q}$ quark model states or gluon-hybrid states can be ruled out. The other approaches consider $Z_c$ states as good candidates for hidden-flavor tetraquark states, molecular states or hadro-charmonia \cite{Olsen:2017bmm, Lebed:2016hpi, Guo:2017jvc, Chen:2016qju, Shepherd:2016dni}. It is also possible, however, that some of the near-threshold peaks can be produced by purely kinematic effects \cite{Swanson:2014tra,Szczepaniak:2015eza,Pilloni:2016obd,Guo:2014iya,Nakamura:2019btl,Nakamura:2019emd}. At the moment, the nature of exotic mesons is still a puzzle in the hadron physics community. More experiments and more detailed theoretical investigations of different reactions are crucial to move towards an understanding of these exotic states.

The reaction $e^+ e^- \to \pi^+ \pi^- \psi(2 S)$ was first measured by the Belle collaboration using the initial state radiation technique \cite{Wang:2014hta,Wang:2007ea}. A clear evidence of a charged intermediate state at $4.05$ GeV was detected in the $\psi \pi^{\pm}$ invariant mass distribution. Recently, the BESIII Collaboration made a high statistics measurement of the same reaction at different $e^+ e^-$ CM energies, $q$ \cite{Ablikim:2017oaf}. At $q=4.416$ GeV, a peak was also observed in the data, which according to an experimental estimate would correspond to a charged charmonium structure with a mass around $4.032$ GeV. However, the total decay width of this new state was not determined due to unresolved discrepancies between the phenomenological fit model and  the data. Moreover, it was noticed that a small variation around $e^+ e^-$ center-of-mass (CM) energy $q= 4.226$ GeV could change significantly the line shape of the invariant mass distributions. This calls for a new analysis that can improve the current description for the Dalitz plot projections for all $e^+ e^-$ CM energies for this process.

In this letter we consider a dispersive approach for this process, which has been successful in the literature in recent years, see for instance Refs. \cite{Chen:2015jgl, Chen:2016mjn, Isken:2017dkw, Chen:2019mgp,Moussallam:2013una} for different applications. The $\pi \pi$ final state interaction (FSI) is accounted for through the Omn\`es formalism which requires the $\pi \pi$ phase shift as input. Two subtraction constants are obtained from the fit to the data. As for the left-hand cuts, we test the data by considering different charged $Z_c$ states in the $t$ and $u$ channels.

The theoretical framework is explained in detail in Sections \ref{Kinematics}, \ref{dispersionR} and \ref{lrc}. Within this approach we fit the experimental invariant mass distributions measured by the BESIII Collaboration \cite{Ablikim:2017oaf} at different $e^+ e^-$ CM energies and show our results in Section \ref{results}. We summarize in Section \ref{summ}.

\section{Kinematics} \label{Kinematics}
The double differential cross section for the process $e^-(p_1)\, e^+(p_2) \to \gamma^* (p_{\gamma^*}) \to  \psi(2S)(p_{\psi})\, \pi^+ (p_{\pi^+})\, \pi^- (p_{\pi^-})$ can be written as
\begin{equation}
\frac{d ^2 \sigma}{d s\, d t} =  \frac{e^2}{2^5 (2 \pi)^3 \, q^6} 
\cdot
\frac{1}{3} \;
\left[
\displaystyle \sum_{\lambda_1 \lambda_2}   
|\mathcal{H}_{\lambda_1 \lambda_2}|^2
\right]
\label{cross-section}
\end{equation}
where $q=\sqrt{p_{\gamma^*}^2}$ and the helicity amplitudes $\mathcal{H}_{\lambda_1 \lambda_2}$ are defined in the usual way,
\begin{align}
\bra{\pi \pi \psi(\lambda_2)} \mathcal{T} \ket{\gamma^{*}(\lambda_1)} = &(2 \pi)^4
\,\delta(p_{\gamma^*} - p_\psi - p_{\pi^+} - p_{\pi^-}) \; \mathcal{H}_{\lambda_1\lambda_2},
\end{align}
with
\begin{align}
\mathcal{H}_{\lambda_1\lambda_2}\equiv &\,\mathcal{H}^{\mu\nu}\epsilon_{\mu}(p_{\gamma^*},\lambda_1)\, \epsilon_{\nu}^{*}(p_{\psi},\lambda_2)\, ,
\end{align} 
and $\lambda_1(\lambda_2)$ denoting the $\gamma^* (\psi(2S))$ helicities respectively.

In the following we choose the Mandelstam variables in terms of the three-body final state,
\begin{align}
s = (p_{\pi^+} + p_{\pi^-})^2, \quad t = (p_{\psi} + p_{\pi^-})^2, \quad u = (p_{\psi} + p_{\pi^+})^2 ,
\end{align}
which satisfy $s + t+ u = q^2 + m_{\psi}^2 + 2 m_{\pi}^2$.
We use the kinematics in the CM frame of the two final pions, and define $z \equiv \cos \theta_s $ as the cosine of the angle between the $p_{\pi^+}$ and the $p_\psi$ momenta,
\begin{align}
t(s, z) = \frac{1}{2}(q^2+m_{\psi}^2 + 2 m_{\pi}^2 - s) + \frac{k(s)}{2}\, z,
\nonumber \\
u(s, z) = \frac{1}{2}(q^2+m_{\psi}^2 + 2 m_{\pi}^2 - s) - \frac{k(s)}{2}\, z,
\end{align}
where 
\begin{align}
k(s) = \frac{1}{s} \sqrt{\lambda(s,q^2,m_{\psi}^2)\, \lambda(s,m_{\pi}^2,m_{\pi}^2)} ,
\end{align}
with $\lambda$
being the K\"allen function. Consequently, $z$ can be written in terms of $t$ and $u$ 
\begin{align}
z = \dfrac{t- u}{k(s)} .
\end{align}

Because of charge conjugation and parity conservation in the process $\gamma^* (1^{--}) \to \psi(2S)(1^{--}) + \pi (0^{-}) + \pi (0^{-})$, the $\pi \pi$-system can only take even values of the total angular momentum $J$ and the isospin values $I_{\pi\pi}=0,2$. Since the photon can only couple to isoscalars or isovectors and the isospin of $\psi(2S)$ is zero, we conclude that only $I_{\pi\pi}=0$ is possible. Under the assumption that left-hand cuts for the reaction with charged and neutral pions are the same, corresponding to dominance of $Z_c$ mechanism, the cross section for $e^+e^- \to \psi(2S) \;  \pi^0 \pi ^0$ differs from the one with the charged pions only by the overall symmetry factor of $1/2$, as it was indeed observed recently in Ref.\,\cite{Ablikim:2017aji}. In the following we omit the isospin index for simplicity, keeping in mind that the transformation coefficient between particle and isospin bases can be absorbed in the overall normalization of the Dalitz plot.

\section{Dispersive Formalism}\label{dispersionR}
In this section we outline a single-channel dispersive formalism to describe the mass distributions for the $e^+ e^- \to \pi^+ \pi^- \psi(2 S)$ process. 
As it will be shown in the next section, the potential kinematic constraints on the helicity amplitudes happen sufficiently far away from the physical region or are very weak so that their impact on the dispersive integral can be ignored. Following the idea of the ``reconstruction theorem" in $\pi\pi$ scattering \cite{Stern:1993rg}, we present the amplitude as a sum of truncated partial wave series in each
of the three channels \cite{Khuri:1960zz, Guo:2014vya, Guo:2015zqa, Pilloni:2016obd}
\begin{align}\label{Eq:isobar}
&\mathcal{H}_{\lambda_1\lambda_2} (s,t,u) \approx  \displaystyle\sum_{J \text{ even}}^{J_{max}} (2J+1) 
\nonumber \\
& \times
\left\lbrace
h _{\lambda_1\lambda_2}^{(J),s}(s)\, d_{\Lambda,0}^{(J)} (\theta_s) 
+ h _{\lambda_1\lambda_2}^{(J),t}(t)\, d_{\Lambda,0}^{(J)}(\theta_t) 
+ h _{\lambda_1\lambda_2}^{(J),u}(u)\, d_{\Lambda,0}^{(J)}(\theta_u)
\right\rbrace ,
\end{align}
where $\Lambda=\lambda_1-\lambda_2$, $d_{\Lambda,0}^{(J)}$ is a Wigner rotation function and $\theta_{s,t,u}$ are the scattering angles in the respective CM frames.
We remark that Eq.(\ref{Eq:isobar}) may be viewed as the most general representation of the constraints imposed by analyticity and crossing symmetry, which is exact for the S- and P-waves, as has been shown in the case of $\pi\pi$ scattering \cite{Stern:1993rg,Knecht:1995tr,Albaladejo:2018gif}.

Truncating the series at $J_{max} = 0$, one can reconstruct the total helicity amplitude as
\begin{align}
\mathcal{H}_{\lambda_1\lambda_2} (s,t,u) =  h_{\lambda_1\lambda_2}^{(0),s}(s) + h_{\lambda_1\lambda_2}^{(0),t}(t) + h_{\lambda_1\lambda_2}^{(0),u}(u)\,.
\end{align}
The partial wave in the $s$-channel can be split as
\begin{align}
&h_{\lambda_1\lambda_2}^{(0)}(s) = \frac{1}{2} \int\limits_{-1}^{+1}dz \,\mathcal{H}_{\lambda_1\lambda_2} (s,t,u) =   h_{\lambda_1\lambda_2}^{(0),s}(s) +  h_{\lambda_1\lambda_2}^{(0),L}(s),\nonumber\\
&h_{\lambda_1\lambda_2}^{(0),L}(s)\equiv \frac{1}{2} \int\limits_{-1}^{+1} dz \; (h_{\lambda_1\lambda_2}^{(0),t}(t) + h_{\lambda_1\lambda_2}^{(0),u}(u))\,,
\end{align}
where the term $h_{\lambda_1\lambda_2}^{(0),L}(s)$ contains the left-hand cuts and the term $ h_{\lambda_1\lambda_2}^{(0),s}(s)$ has only right-hand cuts by definition.

The unitarity equation for the s-channel in the elastic approximation can be written as 
\begin{align}\label{deltaA}
\text{Disc}\, h_{\lambda_1\lambda_2}^{(0)}(s) &\equiv \frac{1}{2\,i}(h_{\lambda_1\lambda_2}^{(0)}(s+i\epsilon) -h_{\lambda_1\lambda_2}^{(0)}(s-i\epsilon))\nonumber\\
&= t^{(0)*}(s)\; \rho(s) \; h_{\lambda_1\lambda_2}^{(0)}(s) \, \theta(s> 4 m_{\pi}^2),
 \end{align}
where $t^{(0)}(s)$ is the S-wave $\pi\pi$ amplitude, and $\rho(s) = \lambda^{1/2}(s,m_\pi^2,m_\pi^2)/s$ is the phase space factor. 
We look for solution in terms of the Omn\`es function
\begin{align}\label{Eq:ansatz}
h_{\lambda_1\lambda_2}^{(0),s}(s) = \Omega^{(0)}(s) \; G^{(0)}_{\lambda_1\lambda_2}(s) ,
\end{align}
which requires as input the $\pi\pi$ phase shift $\delta_{\pi \pi}(s)$
\begin{align}\label{Eq:Omnes}
\Omega^{(0)}(s) =\exp \left[ \frac{s}{\pi} \int\limits_{4m_{\pi}^2}^{\infty} \frac{d s'}{s'} \frac{\delta^{(0)}_{\pi \pi}(s')}{s'-s}\right]\,.
\end{align}
The Omn\`es function satisfies a unitarity relation similar to Eq.\eqref{deltaA},
\begin{align}\label{deltaOmnes}
\text{Disc}\,\Omega^{(0)}(s)= t^{(0)*}(s)\; \rho(s) \; \Omega^{(0)}(s) \, \theta(s> 4 m_{\pi}^2)\,.
\end{align}
Since $\text{Disc}\, h_{\lambda_1\lambda_2}^{(0)}(s)= \text{Disc} \, h_{\lambda_1\lambda_2}^{(0),s}(s)$, one can obtain a dispersion relation for $G^{(0)}_{\lambda_1\lambda_2}$,
\begin{align}
G^{(0)}_{\lambda_1\lambda_2} = - \int\limits_{4m_\pi^2}^{\infty} \dfrac{d s'}{\pi} \dfrac{\text{Disc}\, (\Omega^{(0)})^{-1}(s')\,  h_{\lambda_1\lambda_2}^{(0),L}(s')}{s' - s} .
\end{align}
Consequently, the helicity amplitude with rescattering in the s-channel can be written as
\begin{align}\label{fsi}
&\mathcal{H}_{\lambda_1\lambda_2}(s,t,u) =   h_{\lambda_1\lambda_2}^{(0), t}(t) + h_{\lambda_1\lambda_2}^{(0), u}(u) 
 \\
&+ \Omega^{(0)}(s)\left\lbrace a + b\, s -\dfrac{s^2}{\pi} \int\limits_{4m_\pi^2}^{\infty} \dfrac{d s'}{s'^2} \dfrac{\text{Disc}\, (\Omega^{(0)})^{-1}(s')\,  h_{\lambda_1\lambda_2}^{(0),L}(s')}{s' - s} \right\rbrace \, ,
\nonumber
\end{align}
where we introduced two subtractions (which are functions of photon virtuality $q^2$) in order to reduce the sensitivity to the high energy region and the effects of additional unknown left-hand cuts, such as possible D-meson loops or contact interaction \cite{Chen:2016mjn,Chen:2019mgp}.

\section{Left-Hand Cuts}\label{lrc}
\subsection{Invariant amplitudes and kinematic constraints}
In the general form, the hadron tensor $\mathcal{H}^{\mu \nu}$ can be decomposed into a complete set of Lorenz structures as proposed in Refs. \cite{Tarrach:1975tu,Drechsel:1997xv,Colangelo:2015ama, Danilkin:2018qfn,Danilkin:2017lyn},
\begin{align}
\mathcal{H}^{\mu \nu} = \displaystyle\sum_{i= 1}^{5} F_i L_i^{\mu \nu},
\end{align}
where $F_i$ are the invariant amplitudes and $L_i^{\mu \nu}$ are given by
\begin{align}\label{Li}
L_1^{\mu \nu} =& -p_{\gamma^*}^\nu p_{\psi}^{\mu} + (p_{\gamma^*} \cdot p_{\psi} ) g^{\mu \nu}
\\
L_2^{\mu \nu} =& \left[ 
-\Delta^2 (p_{\gamma^*} \cdot p_{\psi}) + 2 (p_{\gamma^*} \cdot \Delta)  (p_{\psi} \cdot \Delta)
\right] g^{\mu \nu}
+ \Delta^2 p_{\gamma^*}^\nu p_{\psi}^{\mu}
\nonumber \\
&+ 2 (p_{\gamma^*} \cdot p_{\psi})  \Delta^\mu \Delta^\nu 
- 2 (p_{\psi} \cdot \Delta) p_{\gamma^*}^\nu \Delta^\mu 
- 2 (p_{\gamma^*} \cdot \Delta) p_{\psi}^\mu \Delta^\nu 
\nonumber \\
L_3^{\mu \nu} =& (t- u) 
\Big\lbrace
\left[
m_{\psi}^2 (p_{\gamma^*} \cdot \Delta) + q^2 (p_{\psi} \cdot \Delta)
\right] 
\left(
g^{\mu \nu} - \dfrac{p_{\gamma^*}^\nu p_{\psi}^\mu}{(p_{\gamma^*} \cdot p_{\psi}) }
\right) 
\nonumber \\
& +  
\left(
\Delta^{\mu} - \dfrac{(p_{\gamma^*} \cdot \Delta)}{(p_{\gamma^*} \cdot p_{\psi}) } p_{\psi}^\mu
\right)
\left[
-m_{\psi}^2 p_{\gamma^*}^\nu + (p_{\gamma^*} \cdot p_{\psi}) p_{\psi}^\nu
\right]
\nonumber \\ 
&-\left(
\Delta^{\nu} - \dfrac{(p_{\psi} \cdot \Delta)}{(p_{\gamma^*} \cdot p_{\psi}) } p_{\gamma^*}^\nu
\right)
\left[
q^2 p_{\psi}^\mu - (p_{\gamma^*} \cdot p_{\psi}) p_{\gamma^*}^\mu
\right]
\Big\rbrace
\nonumber \\
L_4^{\mu \nu} =& q^2 m_{\psi}^2 g^{\mu \nu} + (p_{\gamma^*} \cdot p_{\psi}) p_{\gamma^*}^\mu p_{\psi}^\nu - q^2 p_{\psi}^\mu p_{\psi}^\nu - m_{\psi}^2 p_{\gamma^*}^\mu p_{\gamma^*}^\nu
\nonumber \\
L_5^{\mu \nu} =& \left( 
q^2 \Delta^\mu - (p_{\gamma^*} \cdot \Delta) p_{\gamma^*}^\mu
\right)
\left( 
m_{\psi}^2 \Delta^\nu - (p_{\psi} \cdot \Delta) p_{\psi}^\nu
\right)\,,\\
\Delta^\mu =& (p_{\pi^+} - p_{\pi^-})^\mu\,.\nonumber
\end{align}
For the S-wave, the $F_{2}$, $F_{3}$ and $F_{5}$ functions vanish, giving rise to simple relations
\begin{align}\label{hpp}
h_{++}^{(0)}(s) &= \frac{s-q^2-m_{\psi}^2}{2} f_1(s) - q^2\, m_{\psi}^2\, f_4(s) ,
\nonumber \\
h_{00}^{(0)}(s) &= - q \, m_{\psi} \left(f_1(s) - \frac{s-q^2-m_{\psi}^2}{2}\, f_4(s) \right),\\
f_i (s)&\equiv  \frac{1}{2} \int\limits_{-1}^{+1} dz \; F_i (s,t).\nonumber
\end{align}
Since invariant amplitudes are free from any kinematic singularities or constraints one can conclude, that the helicity amplitudes are correlated at the kinematic points $s=(q \pm m_\psi)^2$,
\begin{align}\label{KinConstr}
h_{++}^{(0)}(s)\pm h_{00}^{(0)}(s)\sim {\cal O}(s-(q \pm m_\psi)^2)\,.
\end{align}

\subsection{$Z_c$ exchange mechanism}
In the dispersive representation given by Eq.\eqref{fsi}, we approximate the left-hand cut contribution by the exchange of intermediate charmoniumlike charged states in the $t$ and $u$ channels. Based on the experimental data, the mechanism $\gamma^*(q^2) \to \pi + (Z_{c} \to  \psi(2S) + \pi) $ is assumed to be the dominant one. The amplitudes for the process can be written in a general form as,
\begin{equation}
\mathcal{H}^{Z_c}_{\lambda_1\lambda_2} = (V_{Z_{c} \psi \pi})^{\beta \nu} \, 
S_{\nu \mu}(Q_{z}) \,
(V_{\gamma^* \pi Z_{c} })^{\mu \alpha}  \,
\epsilon_{\alpha}(p_{\gamma^*},\lambda_1)\, \epsilon_{\beta}^{*}(p_{\psi},\lambda_2) ,
\label{Mc}
\end{equation}
where $S_{\nu \mu}(Q_z)$ is the axial meson propagator. We adopt the vertex from \cite{Roca:2003uk}, 
\begin{align}
(V_{ Z_{c} \psi \pi})^{\beta \nu} &= C_{Z_{c} \psi \pi} \;  \left[ g^{\beta \nu} \left( p_{\psi} \cdot Q_{z} \right) - p_{\psi}^{\nu} Q_{z}^{\beta} \right] ,
\label{Vz+psipi}
\\
(V_{\gamma^* \pi Z_{c}} )^{\mu \alpha} &= \mathcal{F}_{\gamma^* \pi Z_{c}}(q^2) \; \left[ g^{\alpha \mu} \left( p_{\gamma^*} \cdot Q_{z}\right)  - p_{\gamma^*}^{\mu} Q_{z}^{\alpha} \right] ,
\label{Vyz+}
\end{align}
where $Q_{z} = (p_{\gamma^*} - p_{\pi})$. The form factor $\mathcal{F}_{\gamma^* \pi Z_{c}}(q^2)$ in Eq.\eqref{Vyz+} has a physical meaning only for the on-shell pion and $Z_c$ meson. Below we will consider only $Z_c$ pole contribution, which is well in agreement with Eq.\eqref{Vyz+}. 
In our formalism we will perform an independent fit at each $e^+e^-$ CM energy $q$, without any specific assumptions for $\mathcal{F}_{\gamma^* \pi Z_{c}}(q^2)$. Having enough such energy values, at which one can perform a detailed fit to the data, allows one in principle to reconstruct the line shape of the $e^+e^- \to \psi(2S) \; \pi^+ \pi ^-$ process. In such way one can e.g. test if a description in terms of two Breit-Wigner distributions $Y(4220)$ and $Y(4390)$ as in Ref.\,\cite{Ablikim:2017oaf} is an accurate representation of the cross section.

Due to parity, the helicity amplitudes can be reduced from 9 to 5 independent ones: $\mathcal{H}_{++}$,  $\mathcal{H}_{+-}$,  $\mathcal{H}_{+0}$,  $\mathcal{H}_{0+}$ and  $\mathcal{H}_{00}$.  We observe that $\mathcal{H}_{+-}^{Z_c} =\mathcal{H}_{+0}^{Z_c} = \mathcal{H}_{0+}^{Z_c} \approx 0$ compared to $\mathcal{H}_{++}^{Z_c}$ and $\mathcal{H}_{00}^{Z_c}$, which confirms our assumption that the process is dominated by the S-wave. Also, for our particular kinematics the approximation $|\mathcal{H}_{++}^{Z_c}| \approx |\mathcal{H}_{00}^{Z_c}|$ can be made with less than $1\%$ error in the physical region (similar observation was also made in Refs. \cite{Chen:2015jgl,Chen:2016mjn}). Therefore,
\begin{align}
\displaystyle\sum_{\lambda_1\lambda_2}  |\mathcal{H}_{\lambda_1\lambda_2}^{Z_c}|^2  \approx 2 \,|\mathcal{H}_{++}^{Z_c}|^2 + |\mathcal{H}_{00}^{Z_c}|^2 \approx 3\, |\mathcal{H}_{++}^{Z_c}|^2,
\label{Hsquared}
\end{align}
and we can ignore the effects of the kinematical constraints given by Eq.\eqref{KinConstr}.

Using the helicity amplitudes calculated via Eq.\eqref{Mc}, we show the invariant amplitudes $F_i^{Z_c}(s,t)$ that give the dominant contribution for the S-wave
\begin{align}\label{Fs}
F_1^{Z_c} (s, t) =& - \frac{\mathcal{F}_{\gamma^* \pi Z }(q^2)\, C_{Z \psi \pi} }{8} \left(
\dfrac{4\,t + q^2 + m_{\psi}^2}{t - m_z^2} + \dfrac{4\,u + q^2 + m_{\psi}^2}{u - m_z^2}
\right) ,
\nonumber \\
F_4^{Z_c} (s, t) =& - \frac{\mathcal{F}_{\gamma^* \pi Z }(q^2)\, C_{Z \psi \pi} }{4} \left(
\dfrac{1}{t - m_z^2} + \dfrac{1}{u - m_z^2}
\right).
\end{align}
Due to the polynomial behavior of the amplitudes at high energies, we will consider the pole contribution, which corresponds to fixing $t= m_z^2$ and $u= m_z^2$ in the numerators. This procedure is in line with the definition of the transition form factor $\mathcal{F}_{\gamma^* \pi Z_{c}}(q^2)$ and almost does not change the amplitude in the physical region. 

\begin{figure}[t]
\centering
\includegraphics[width =0.40\textwidth]{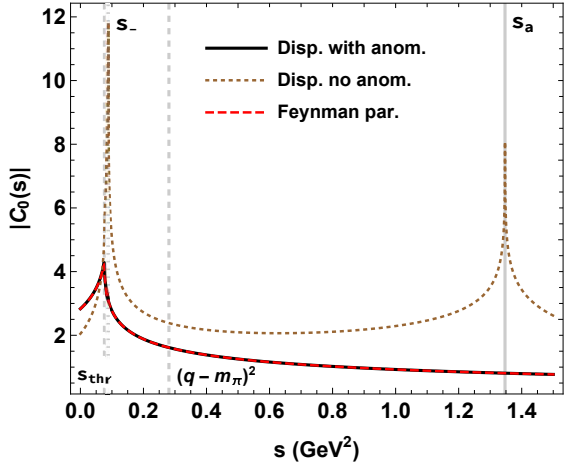}
\caption{Comparison of the absolute values of the scalar triangle loop function $C_0(q^2,m_\psi^2,s,m_\pi^2,m_z^2,m_\pi^2)$ calculated numerically using Feynman parameters (dashed red line) and dispersively with (solid black line) or without (dotted brown line) anomaly piece given in Eq.(\ref{L(s)new}). The result is illustrated for $q=4.226$ GeV and $Z_c(3900)$ as intermediate state.}
\label{anom-comp}
\end{figure}

\subsection{Anomalous threshold} 
Depending on the kinematics of the reaction, left- and right- hand cuts may overlap leading to an additional, anomalous piece in the dispersive integral of Eq.\eqref{fsi}. The left-hand branch points of partial wave amplitudes Eq.\eqref{hpp} can be determined by the endpoint singularities of the $t$- and $u$-channel projection integrals
\begin{align}
L(s) &\equiv 
\int\limits_{-1}^{+1} 
\dfrac{d z}{t - m_z^2}
=
\int\limits_{-1}^{+1} \dfrac{d z}{u - m_z^2} 
= - \frac{2}{k(s)} \log \left( \frac{\chi(s) + 1}{\chi(s) - 1}\right) ,\nonumber\\
&\chi(s) = \dfrac{2 m_z^2 - q^2-m_{\psi}^2 - 2 m_{\pi}^2 + s}{k(s)}\,,
\label{L(s)}
\end{align}
and are given by
\begin{align}\nonumber
s_{\pm} =& \frac{1}{2} \left[ 
q^2+m_{\psi}^2 + 2 m_{\pi}^2 - m_z^2 - \dfrac{(q^2-m_\pi^2)(m_\psi^2-m_\pi^2)}{m_z^2}
\right] 
\pm \dfrac{k(m_z^2)}{2\, m_z^2}.
\end{align}
When $q^2>2m_\pi^2 + 2m_z^2 - m_\psi^2$, the branch point $s_-$ moves from the unphysical (square-root) Riemann sheet onto the physical sheet and requires the proper deformation of the integration contour \cite{Karplus:1958zz,Mandelstam:1960zz,Hoferichter:2013ama} (see also \cite{Lutz:2015lca} where a general spectral representation is established for the arbitrary masses case). Effectively, it corresponds to including an additional piece to Eq.(\ref{L(s)}), which is related to the discontinuity of $L(s)$ on the anomalous cut
\begin{align}\label{L(s)new}
&L(s) \to L(s) \, \underbrace{- i \frac{4\,\pi}{k(s)}\, \theta(s_{-} < s < s_{a})}_{\text{anomalous piece}}\,,\\
&s_{a}=2m_\pi^2 +m_{\psi}^2+q^2 - 2 m_z^2\,,\nonumber
\end{align}
and making the analytical continuation $q^2\to q^2+i\epsilon$ \cite{Moussallam:2013una,Bronzan:1963mby}. The location of $s_{a}$ is determined by the condition that the imaginary part of $L(s)$ changes sign. To cross-check whether this prescription is correct, we consider a toy model of scalar fields and calculate a triangle loop function. In Fig.\,\ref{anom-comp} two results are shown: the direct calculation via Feynman parameters and the result of a dispersive representation. The exact agreement is achieved only when the anomalous piece in Eq.\eqref{L(s)new} is taken into account.

We note, that the considered $e^+e^-$ center-of-mass energies satisfy the condition $q^2>(m_Z+m_\pi)^2$ and also $m_Z^2 > (m_\psi + m_\pi)^2$. It implies that $Z_c$ can be produced on-shell and it calls for taking into account the width of $Z_c$ in the rescattering (dispersive) part. The proper implementation requires modeling the propagator using a spectral representation, i.e., it should have sound analyticity properties, such as pole on the unphysical Riemann sheet and the right-hand cuts starting at $\pi J/\psi$ and $D\bar{D}^*$ thresholds. This analysis is beyond the scope of our paper due to the lack of experimental information. We checked, however, on the example of the toy model that a naive implementation of the finite width hardly affects the results of dispersive integral due the narrowness of $Z_c$. Therefore, for the rescattering part, we neglect the width of $Z_c$, while in the evaluation of the first two terms of Eq.\eqref{fsi}, we include the finite width of $Z_c$ to Eq.(\ref{Fs}).

\begin{figure}[t]
\includegraphics[width =0.235\textwidth]{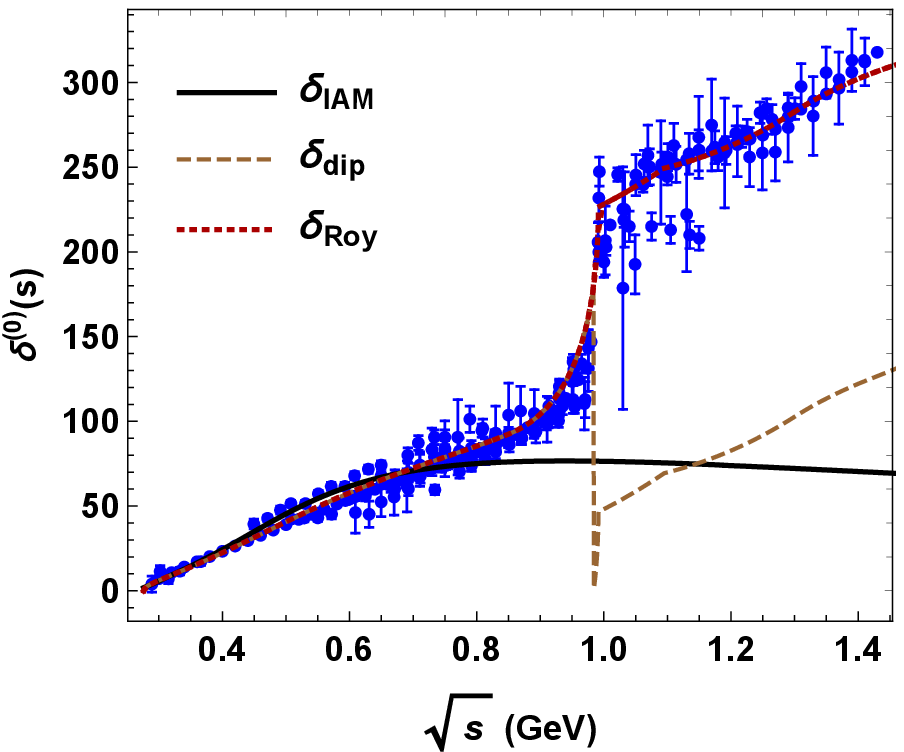}
\includegraphics[width =0.235\textwidth]{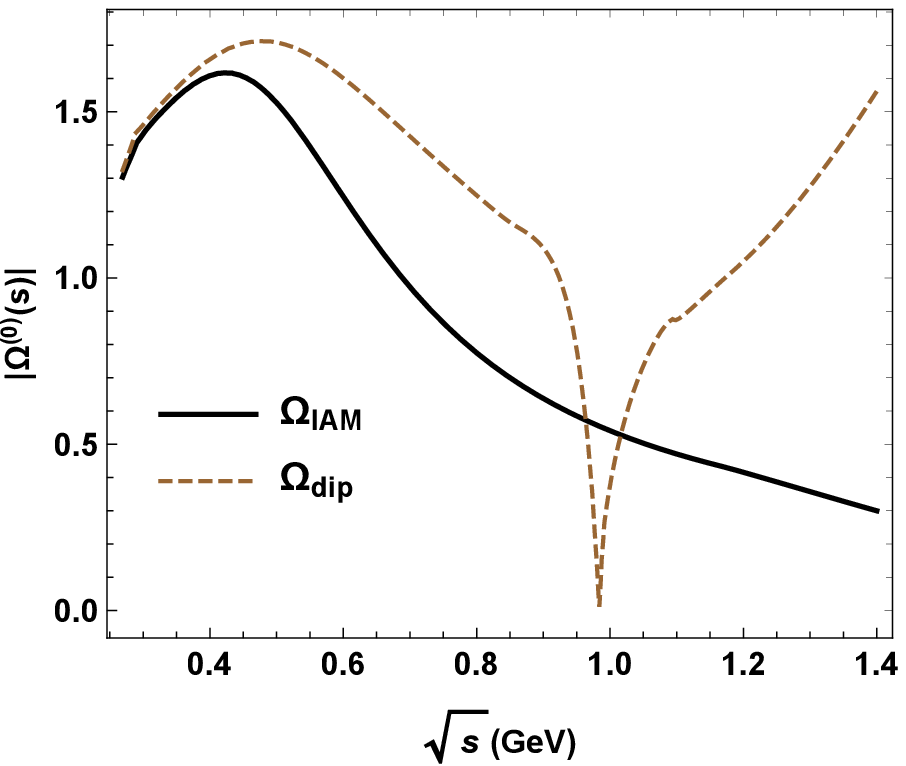}
\caption{The left panel shows the experimental data on the $\pi \pi$ phase shift for the S-wave, $I=0$ in comparison with the Roy equation analysis (red dotted line) \cite{GarciaMartin:2011cn}, the modified phase shift which exhibit a sharp “dip” like behaviour at the point where phase shift crosses $\pi$ (dashed brown line) \cite{Moussallam:2013una,Oller:2007sh} and the phase shift from the modified IAM (solid black line) \cite{GomezNicola:2007qj}. The respective modulus of the Omn\`es functions are shown on the right panel for $\delta_{\text{dip}}$ and $\delta_{\text{IAM}}$.}
\label{Omnes-curves}
\end{figure}

\begin{table*}[t]
\renewcommand*{\arraystretch}{1.4}
\begin{tabular*}{\textwidth}{@{\extracolsep{\fill}}|c|cclccc|@{}}
\hline 
$q$ (GeV) &  $|e\, \mathcal{F}_{\gamma^* \pi Z_c }\, C_{Z_c \psi \pi} |^2$ (GeV$^{-4}$) & $|a /\mathcal{F}_{\gamma^* \pi Z_c }\,C_{Z_c \psi \pi}|$ (GeV$^{2}$) & $\phi_{a}$ (rad) & $|b/\mathcal{F}_{\gamma^* \pi Z_c }\,C_{Z_c \psi \pi}|$ & $\phi_{b}$ (rad)& $\chi_{\text{red}}^2$   \\
\hline 
4.226  & $3.7 (5) \cdot 10^{-6}$ & $ 1.09(25) \cdot 10^{3}$ & $5.60(20)$ & $ 7.4(16) \cdot 10^{3}$ & $1.98(14)$ & 1.16 \\
4.258  & $1.3(3) \cdot 10^{-6}$ & $0.15 \cdot b$ & $2.61(25)$ & $8.2(14) \cdot 10^{3}$ & $5.40(19)$ & 1.01 \\
4.416  & $2.0(4) \cdot 10^{-6}$ & $2.02(24)\cdot 10^{3}$ & $2.28(18)$ & $ 9.5(10) \cdot 10^{3}$ & $5.57(15)$ & 1.38 \\
\hline
\hline
$q$ (GeV) &  $|e\,\mathcal{F}_{\gamma^* \pi Z_c }\,C_{Z_c \psi \pi} |^2$ (GeV$^{-4}$) & $a /b$ (GeV$^{2}$) & $\phi_{a}$ (rad) & $|b|^2$ (GeV$^{-4}$)  & $\phi_{b}$ (rad) & $\chi_{\text{red}}^2$  \\
\hline
4.358 & -& $-0.165(4)$ & - & $ 3.59(19) \cdot 10^{2}$ & -  & 0.83\\ 
\hline
\end{tabular*}
\caption{Fit parameters entering Eq.\eqref{fsi} for four different $e^+e^-$-CM energies $q$.\label{fits-results}}
\end{table*}

\subsection{$\pi\pi$ rescattering}
To compute the integral in Eq.\eqref{fsi}, we need to specify the input for the Omn\`es function in Eq.\eqref{Eq:Omnes} that encodes the $\pi\pi$ rescattering part. In this work, we only consider elastic unitarity, which is essentially exact in the physical regions of all Dalitz plot projections. The phase shift $\delta^{(0)}_{\pi\pi}(s)$ we extract from a single-channel modified inverse amplitude method (mIAM) \cite{GomezNicola:2007qj}, similar to \cite{Colangelo:2017qdm,Colangelo:2017fiz}. The benefit of this approach is twofold. First, it reproduces the $f_0(500)$ parameters (such as pole and coupling) consistent with the Roy equation solutions \cite{GarciaMartin:2011cn}. Second, there is no sharp onset of $K\bar{K}$ inelasticity due to the $f_0(980)$ resonance. The latter requires a coupled-channel treatment with inclusion of $K\bar{K}$ intermediate states. Alternatively to the input from the mIAM, in the elastic approximation one can construct a modified Omn\`es function with a phase which exhibits a sharp "dip" behaviour at two-kaon threshold \cite{Moussallam:2013una,Oller:2007sh}. The impact on the Omn\`es function is shown in Fig.\,\ref{Omnes-curves}. We observe that both approaches give similar results only at very low energies. Whereas at larger energies, the solution based on the "dip" like phase shift exhibit a cusp across the inelastic region, while Omn\`es function based on the mIAM phase shift is completely smooth. We checked that, given the number of subtractions we are using, both solutions lead to equivalent results for the Dalitz plot projections fits, however, we find the mIAM input to be more suitable for the dispersive formalism with elastic unitarity.

\section{Results and Discussion} \label{results}

In the previous section, we described our theoretical approach, which consists in using a charged exotic state as an intermediate particle and the dispersion theory to account for the two-pion final state interaction, as shown in Eq.\eqref{fsi}. 
With that, we perform a simultaneous fit of the experimental invariant mass distributions $M_{\psi \pi^{\pm}}^2$ and $M_{\pi^+ \pi^{-}}^2$ at different $e^+e^-$-CM energies $q=4.226;\, 4.258; \, 4.358; \, 4.416$ GeV. 
From the total cross section normalization, as given in Ref.\,\cite{Ablikim:2017oaf}, we extract the normalized mass distributions by assuming a constant detector efficiency.

For each energy $q$ we consider initially two complex subtraction constants, $|a|, \phi_a $ and $ |b|, \phi_b$ respectively, and a global normalization, which contains the product of the coupling constants $\mathcal{F}_{\gamma^* \pi Z } C_{Z \psi \pi}$.
The subtraction constants are complex due to the specific analytic structure of the $Z_c$ exchange left-hand cut which overlaps with the unitarity cut (see Eq.(\ref{L(s)})). All the fit parameters are supposed to depend on $q$. However, for nearby values of $q$ we do not expect a large variation in the parameter values. Despite using the same expression to fit the data, the parameter values are completely driven by the experimental distribution, which exhibits different features for each $e^+e^-$-CM energies $q$. The results of the fits are shown in Tables \ref{fits-results} and \ref{par-comp}. 

\begin{figure*}[h!]
\scriptsize
\centering
\includegraphics[height=4.75cm]{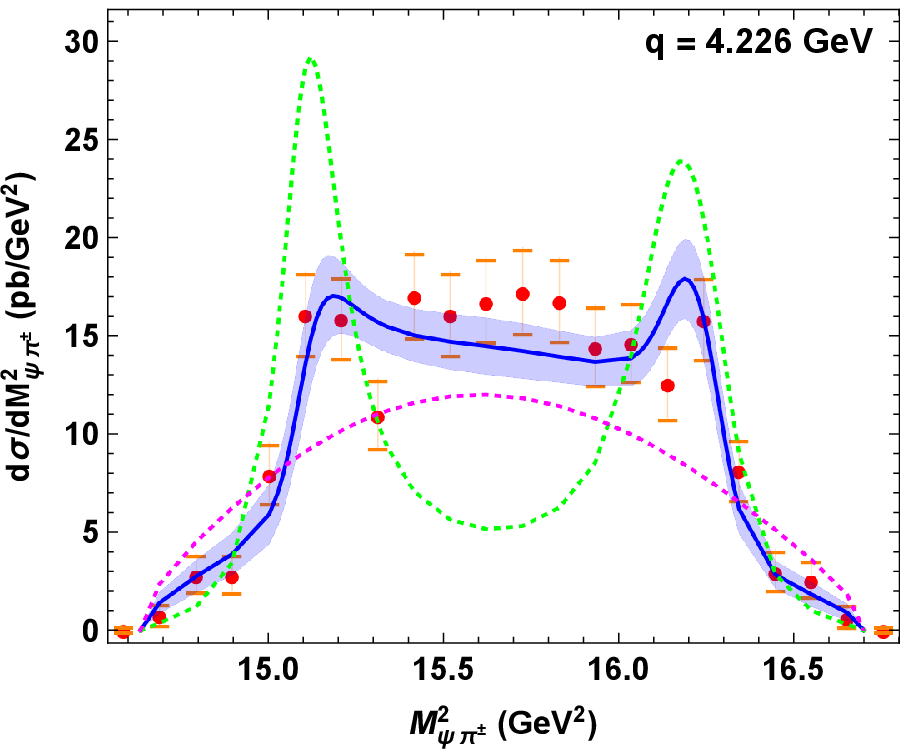}\qquad\includegraphics[height=4.75cm]{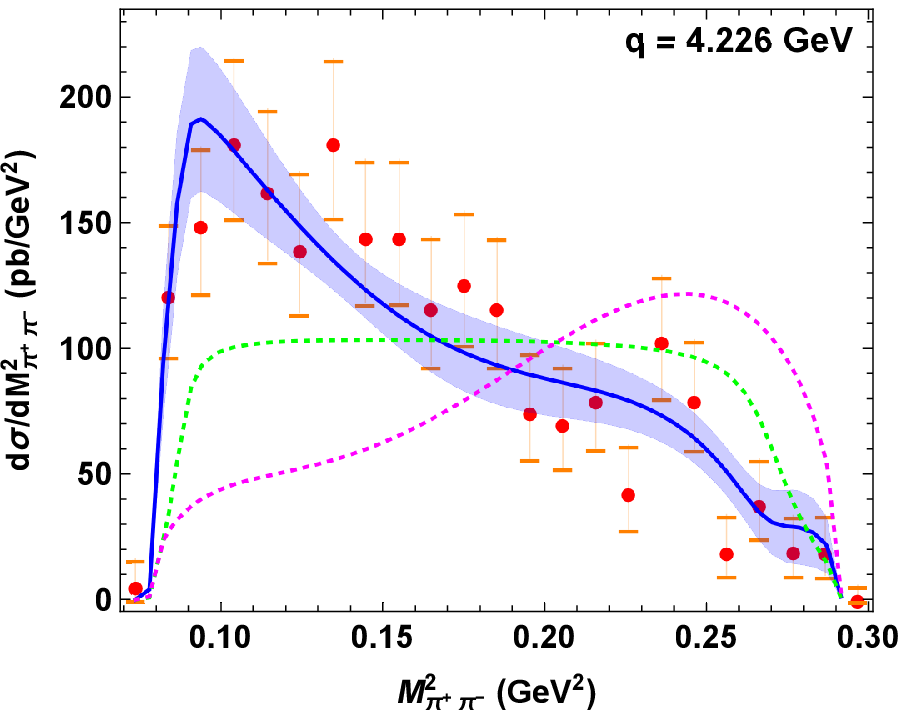}\\
\includegraphics[height=4.75cm]{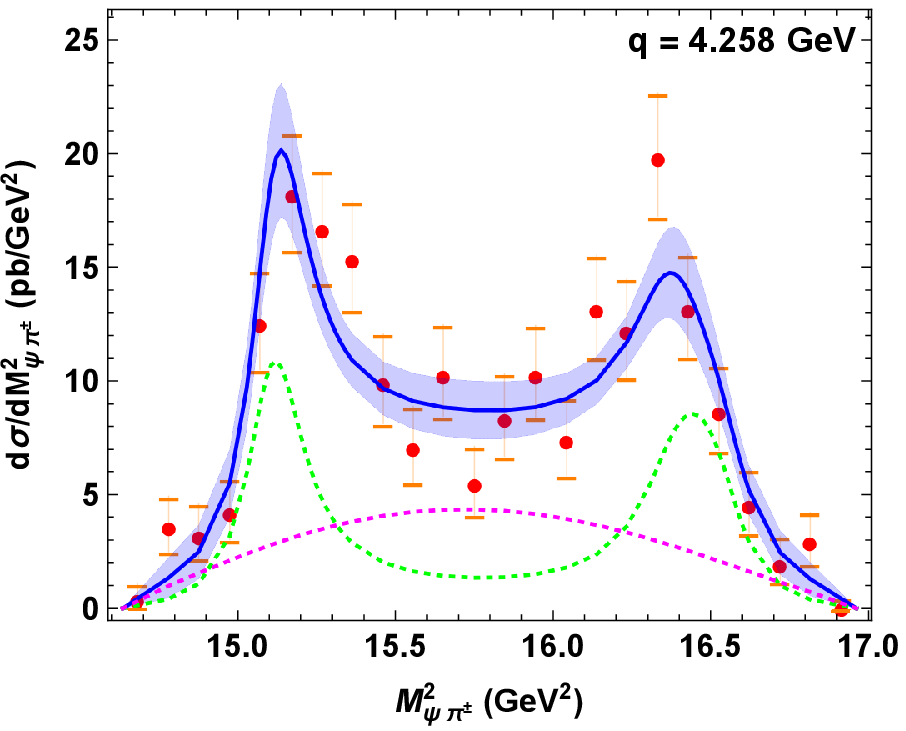}\qquad\includegraphics[height=4.75cm]{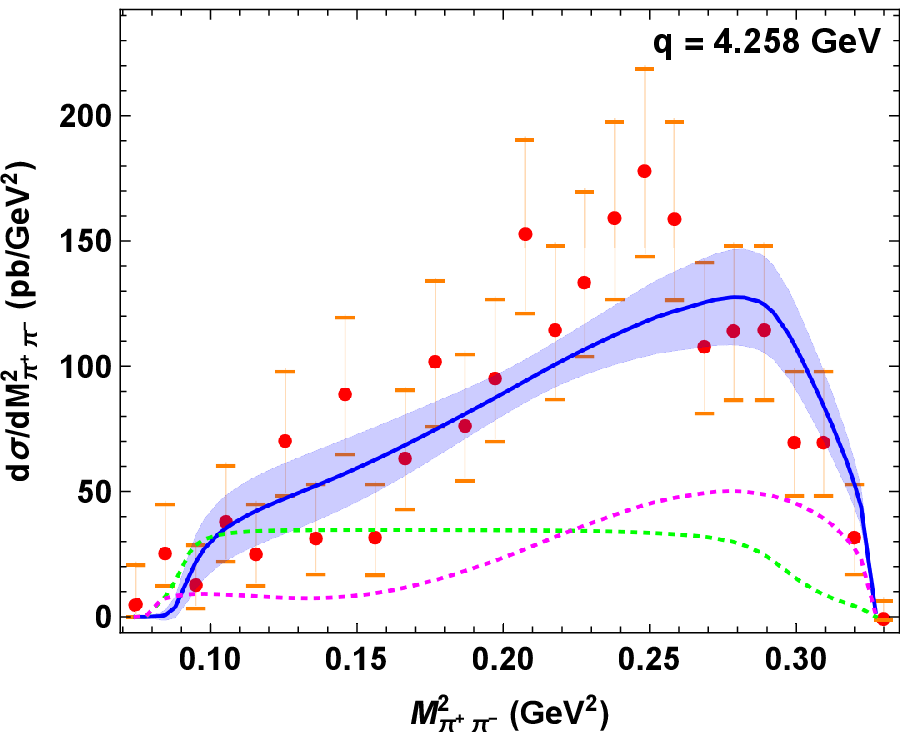}
\caption{Fits for $q=4.226$ GeV and $4.258$ GeV including the intermediate state $Z_c(3900)$. The red dots are the data points from BESIII \cite{Ablikim:2017oaf} normalized as explained in the text. The blue curves are the fit performed with two subtractions, where the purple bands are calculated by propagating the statistical error of the parameters with $95 \%$ confidence level. The green dotted curves are the contribution of only the intermediate state $Z_c(3900)$ and the magenta dotted curves are the contribution from the $\pi \pi$-FSI (i.e. when $C_{Z_c\psi\pi}=0$).\label{Zc3900}}
\end{figure*}

\begin{figure*}[h!]
\centering
\includegraphics[height=4.75cm]{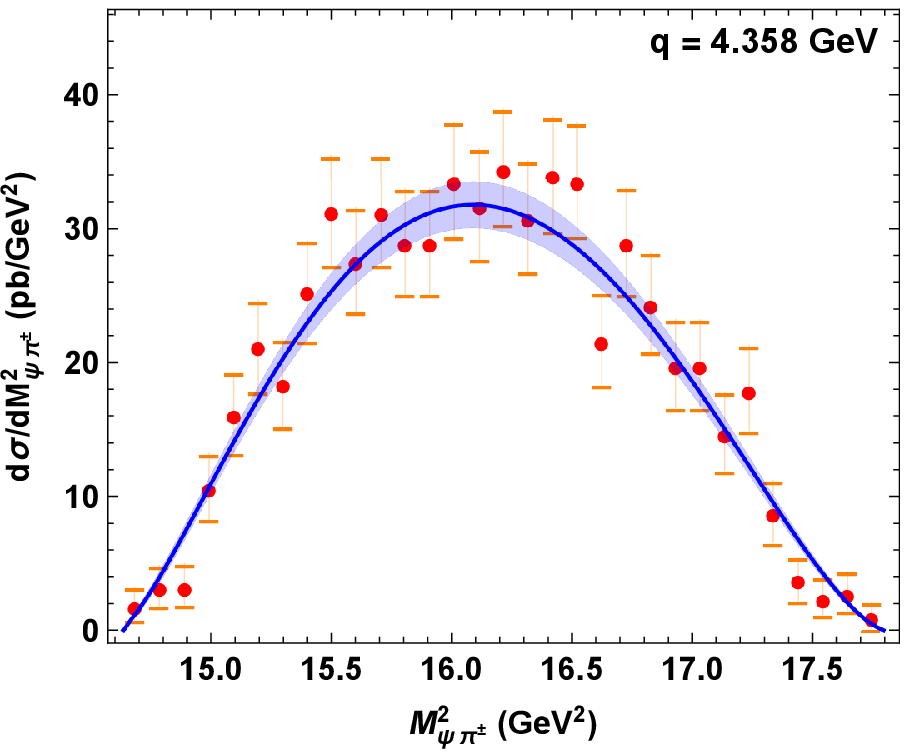}\qquad\includegraphics[height=4.75cm]{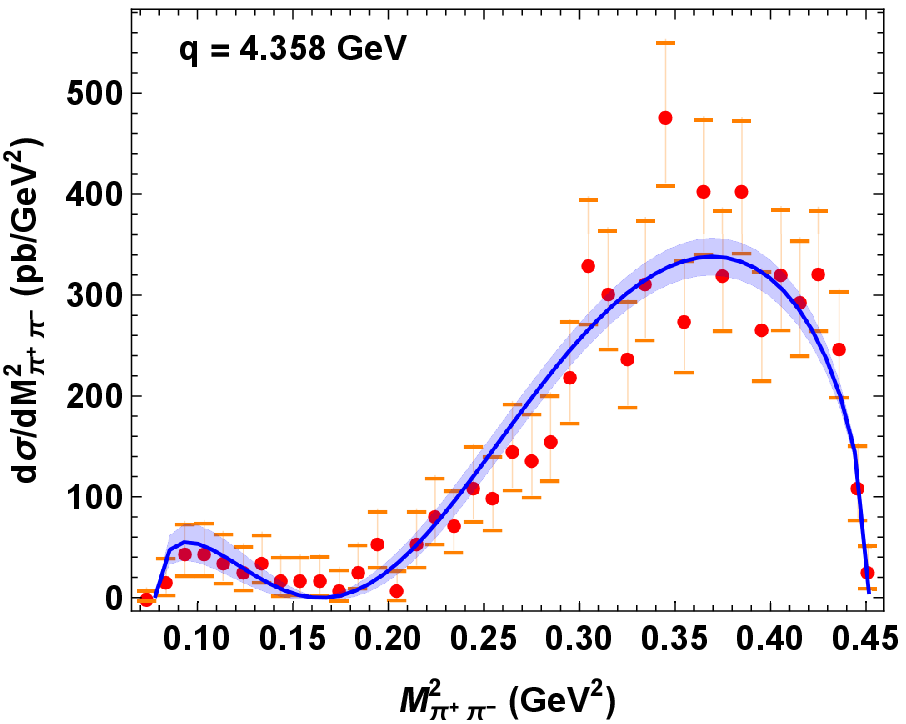}
\caption{Fit for $q= 4.358$ GeV without $Z_c$ intermediate state, i.e. considering the contribution from the $\pi \pi$-FSI.\label{q4358}}
\end{figure*}

At $q=4.226$ GeV we achieve a very good description of the experimental data for both invariant mass distributions, considering the already established $Z_c(3900)$ as the intermediate state, with $m_{Z_c}= 3.8866$ GeV and $\Gamma_{Z_c}= 28.1$ MeV from Ref.\,\cite{Tanabashi:2018oca}. As one can see in Fig.\,\ref{Zc3900}, this result is an improvement over the phenomenological description in Ref.\,\cite{Ablikim:2017oaf}, where the $M_{\psi \pi}^{2}$ and $M_{\pi \pi}^{2}$ mass distributions could not be fitted simultaneously. For $q=4.258$ GeV we consider the same assumptions as for the $q=4.226$ GeV case and also obtain a good description of the data. However, the fit is not sensitive to the value of the first subtraction constant $a$. Therefore, we fix $a(q=4.258)$ by constraining the ratio of the subtraction constants to be the same as in the lower $q=4.226$ GeV, i.e. $a/b_{q=4.258}=a/b_{q=4.226}$ and obtain an excellent $\chi_{\text{red}}^2\equiv\chi^2/N_{\text{d.o.f}}= 1.01$. At $q=4.358$ GeV we observe that the best fit does not require an intermediate $Z_c$ state in the left-hand cuts. The fit with real values for two subtraction constants multiplied by the $\pi \pi$ Omnes function perfectly describe the data, as shown in Fig.\,\ref{q4358}. In other words, this implies that for $q=4.358$ GeV the left-hand cuts are dominated by the contact interaction which are absorbed in the subtraction constants in the present framework.

\begin{figure}[h!]
\centering
\includegraphics[width=8cm]{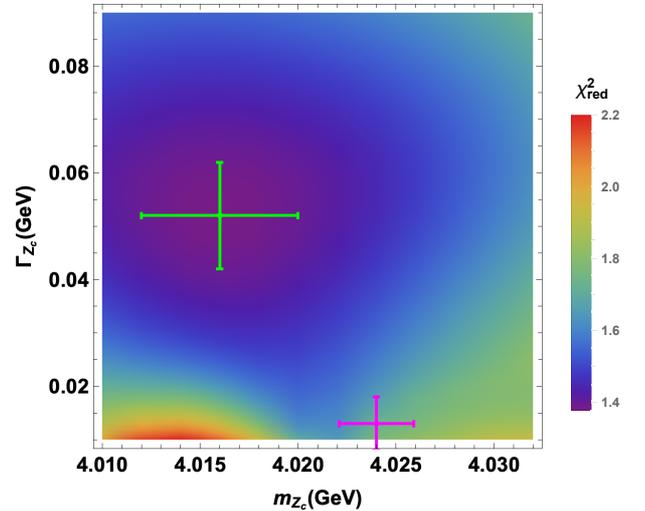}
\caption{$\chi^2_{\text{red}}$ as a function of the mass and width of the heavier intermediate $Z_c$ state at $q=4.416$ MeV. The minimum of $\chi^2_{\text{red}}$ is indicated by a green cross, with its estimated uncertainty. The PDG average of $Z_c(4020)$ state \cite{Tanabashi:2018oca} is indicated by a magenta cross.}
\label{densityplot}
\end{figure}

\begin{figure*}[h!]
\centering
\includegraphics[height=4.75cm]{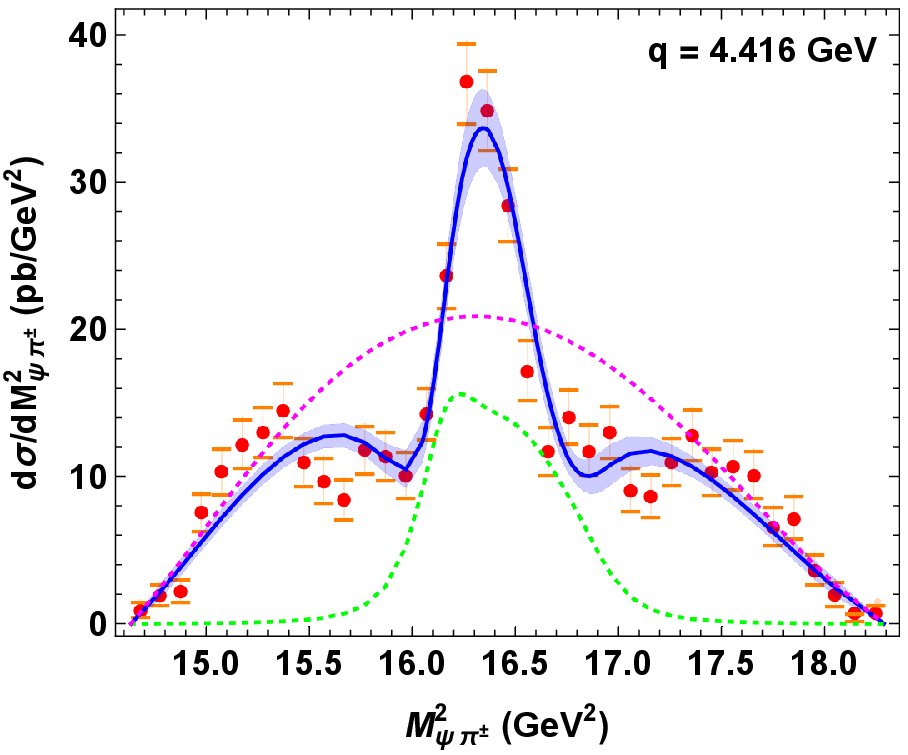}\qquad\includegraphics[height=4.75cm]{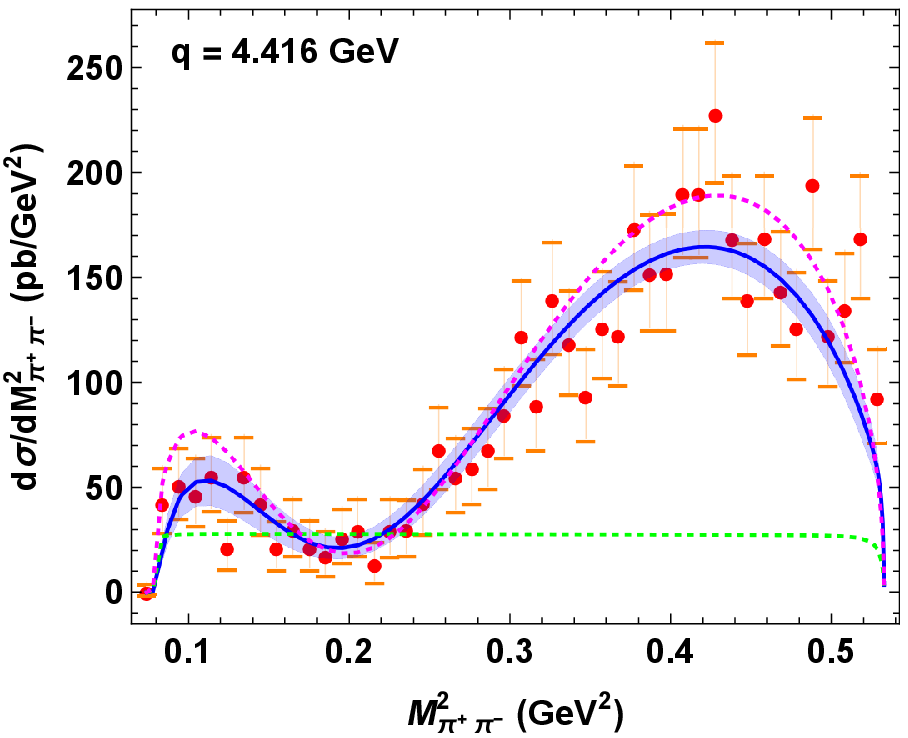}
\caption{Fit for $q=4.416$ GeV including the heavier $Z_c$ intermediate state.\label{Zc4030}}
\end{figure*}

For $q=4.416$ GeV, we test the experimental claim of a possible observation of a heavier charged intermediate state \cite{Ablikim:2017oaf}. Its parameters were not well established due to unresolved discrepancies between a model fit and the data. In Fig.\,\ref{densityplot}, we analyze the dependence of the $\chi^2_{\text{red}}$ on the mass and the width of the possible heavier $Z_c$ state. For the best $\chi^2_{\text{red}}$ we obtain an accurate description of the pronounced enhancement in the data (see Fig.\,\ref{Zc4030}) for the mass $m_{Z_c} = 4.016(4)$ GeV and the width $\Gamma_{Z_c} = 52(10)$ MeV. However, we notice that the $\chi^2_{\text{red}}$ distribution is wide and smooth. Therefore, we cannot completely rule out that the signal seen at this energy corresponds to $Z_c(4020)$ ($m = 4.024(2)$ GeV and $\Gamma = 13(5)$ MeV according to PDG \cite{Tanabashi:2018oca}) observed in the reactions $e^+ e^- \to D^* \bar{D}^* \pi$ \cite{Ablikim:2014dxl,Ablikim:2015vvn} and $e^+ e^- \to h_c \pi \pi$ \cite{Ablikim:2013wzq,Ablikim:2013emm}. 

The invariant mass distributions of the neutral counterpart $e^+ e^- \to \psi(2S) \pi^0 \pi^0$ at the same $e^+e^-$-CM energies, were measured experimentally in Ref.\,\cite{Ablikim:2017aji}. As we pointed out above, due to isospin symmetry the cross section for $e^+e^- \to \psi(2S) \;  \pi^0 \pi ^0$ differs from the one with the charged pions only by the overall symmetry factor of $1/2$. However, we do not include this data in our fits since it has lower statistics and does not bring additional constraints on the fit. Future larger statistical samples are desired for both charged and neutral decay channels, to investigate how the already established $Z_c(3900)$ state contributes to $q=4.416$ GeV.

\begin{table}[h!]
\renewcommand*{\arraystretch}{1.4}
\begin{tabular*}{\columnwidth}{@{\extracolsep{\fill}}|c| c l c c|}
\hline  
$q$ (GeV) & $|a /b|$ (GeV$^{2}$) & $\phi_{a}/\pi$ & $\phi_{b}/\pi$ & $\chi_{\text{red}}^2$ \\
\hline 
\hline 
4.226 &  $0.15(5)$ & $1.78(6)$ & $0.63(4)$ & 1.16 \\
4.258 & $0.15$ & $0.83(8)$ &  $1.71(6)$ & 1.01 \\
4.358 & $0.165(4)$ & - & - & 0.83 \\
4.416 & $0.21(3)$ & $0.72(6)$ & $1.77(5)$ & 1.38 \\
\hline 
\end{tabular*}
\caption{Comparison of the complex subtraction constants of Eq.\eqref{fsi} $a= |a|e^{i\phi_a}$ and $|b|e^{i\phi_b}$ for four different $e^+e^-$-CM energies $q$.\label{par-comp}}
\end{table}

\section{Summary} \label{summ}
In this letter, we presented an amplitude analysis of the reaction $e^+ e^- \to \psi(2S)\, \pi^+ \pi^-$ at different $e^+ e^-$-CM energies $q$. The final state interaction of the two pions is treated using the dispersion theory and we studied quantitatively the contribution of the charged exotic mesons as intermediate states. We observed that the $Z_c(3900)$ state plays an important role to explain the invariant mass distribution at both $q=4.226$ and $q=4.258$ GeV. To explain the sharp narrow structure at $q=4.416$ GeV, a heavier charged state is needed instead, with $m_{Z_c} = 4.016(4)$ GeV and $\Gamma_{Z_c} = 52(10)$ MeV. The latter is not necessarily a new state since its mass is compatible with the already known $Z_c(4020)$. For $q=4.358$ GeV no intermediate $Z_c$ state is necessary for left-hand cuts in order to describe both $\psi\pi$ and $\pi\pi$ line shapes. It points to another left-hand contribution which we absorbed in the subtraction constants. We also conclude that the $\pi\pi$-FSI is the main mechanism to describe the $\pi\pi$ invariant mass distribution for all four $e^+ e^-$-CM energies.

\section*{Acknowledgements}
The authors acknowledge Zhiqing Liu and Achim Denig for useful discussions about the experimental data. D.A.S.M. also thanks Matthias Heller for helping to access the supercomputer Mogon at Johannes Gutenberg University Mainz.
This work was supported by the Deutsche Forschungsgemeinschaft (DFG, German Research Foundation), in part through the Collaborative Research Center [The Low-Energy Frontier of the Standard Model, Projektnummer 204404729 - SFB 1044], and in part through the Cluster of Excellence [Precision Physics, Fundamental Interactions, and Structure of Matter] (PRISMA$^+$ EXC 2118/1) within the German Excellence Strategy (Project ID 39083149).

\bibliographystyle{apsrevM}
\bibliography{Daniel_bibliography}

\end{document}